\documentclass[pre,aps,twocolumn,showpacs,preprintnumbers,amsmath,amssymb]{revtex4-1}

\usepackage{esint} 
\let\oldoiint\oiint
\renewcommand{\oiint}{\oldoiint\limits}

\usepackage{graphicx}
\usepackage{bm}
\newcommand \be{\begin{eqnarray}}
\newcommand \ee{\end{eqnarray}}
\newcommand \ba{\begin{align}}
\newcommand \eea{\end{align}}

\newcommand \V{\vec}

\begin{document}
\title{Induced voltage in an open wire}
\author{K. Morawetz$^{1,2,3}$, M. Gilbert$^1$, A. Trupp$^4$}
\affiliation{$^1$M\"unster University of Applied Sciences,
Stegerwaldstrasse 39, 48565 Steinfurt, Germany}
\affiliation{$^2$International Institute of Physics (IIP)
Federal University of Rio Grande do Norte
Av. Odilon Gomes de Lima 1722, 59078-400 Natal, Brazil
}
\affiliation{$^{3}$ Max-Planck-Institute for the Physics of Complex Systems, 01187 Dresden, Germany}
\affiliation{$^{4}$ Brandenburg 
University of Applied 
Police 
Sciences, Bernauer Stra\ss{}e 146,
16515 Oranienburg, Germany}
\begin{abstract}
A puzzle arising from Faraday's law is considered and solved concerning the
question which voltage is induced in an open wire with a time-varying
homogeneous magnetic field. In contrast to closed wires where the voltage is determined by the time variance of magnetic field and enclosed area, in an open wire we have to integrate the electric field along the wire. It is found that the longitudinal electric field contributes with 1/3 and
the transverse field with 2/3 to the induced voltage. In order to find the electric fields the sources of the magnetic fields are necessary to know. The representation of a homogeneous and time-varying magnetic field implies unavoidably a certain symmetry point or symmetry line which depend on the geometry of the source. As a consequence the induced voltage of an open wire is found to be the area covered with respect to this symmetry line or point perpendicular to the magnetic field. This in turn allows to find the symmetry points of a magnetic field source by measuring the voltage of an open wire placed with different angles in the magnetic field.  We present exactly solvable models for a symmetry point and for a symmetry line, respectively. The results are applicable to open circuit problems like corrosion and for astrophysical applications.
\end{abstract}
\maketitle

\section{Introduction}


Faraday's law is standard textbook knowledge. The induced voltage of a closed circle in a magnetic field is either caused by the time-dependent change of the enclosed area or the time-dependent change of the magnetic field \cite{GIKLY06,Gr12}.  Interesting puzzles and the induction in deformable circuits can be found in \cite{Sco95}. Faraday's induction experiments now have gained a certain revival when nanostructures are considered \cite{KHHA08} and play a crucial role in type II superconductors \cite{KL72,LR06}, see \cite{LKMBY07} for references. The magnetic field effects in currents are even used for measuring the speed of light \cite{Sp12}.

Though induction in closed wires and the forces acting on wires in magnetic fields \cite{Re11} are well understood, the induction in open wires is rarely studied, probably since the effects there are especially puzzling. Though magnetic effects due to open circuits have been known for more than 100 years they are still of interest with respect to corroding problems in ferromagnetic electrodes. For an overview of the experimental activities and their history see \cite{DCGL05}. 
Since most experiments are performed with respect to the question of corroding materials \cite{W01,SI02,Su10}, it is overlaid by the problem of chemical reactions. Then non-equilibrium situations have to be considered such that Lorentz and gradient forces become important on a stream in anodic dissolution of microstructures \cite{Bu05}. These magnetic field effects are crucial for patterning of electrodeposits \cite{Du12}. Also eddy currents, measured e.g. with contact-less methods \cite{Kr99}, are still a hard problem. In \cite{DCGL05,Su10} the orientation of the electrode in the magnetic field
reveals opposite responses when oriented parallel or perpendicular to the field. This will be explained by our approach. 

Despite these variety of applications the simple question what voltage is induced in an open wire or circuit when placed in a homogeneous and time-varying magnetic field is not treated to the best of our knowledge. Here we want to resolve this puzzle providing a unique expression of the voltage induced at the ends of an open wire within time-dependent and homogeneous magnetic fields and its dependence on the direction of the magnetic field. 

\subsubsection{Paradox}
\begin{figure}[h]
\includegraphics[width=5cm]{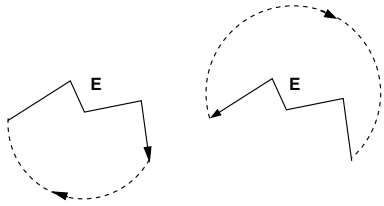}
\caption{Induction in two identical wires (solid) differing
only by two closing paths (dashed) leading to an opposite sign of the induced voltage.}
\label{wider}
\end{figure}

A first paradox arises if asking the simple question which voltage might be induced in an open curved wire exposed to a time-varying but homogeneous magnetic field. 
A gedanken experiment seems to convince us that this voltage is undetermined. Dependent whether one closes the wire clockwise or anticlockwise, a different sign of the induced voltages is obtained at its ends as shown in Fig. \ref{wider}. The path used to turn theses wires into a closed loop will either induce an electric field from left to right or from right to left.
Many different setups can be constructed which show the same contradiction. 

The solution of this paradox is enlightening the ingeniousness of Faraday's law. In order to {\em measure} the voltage one has to close the open wire in {\em some way} which provides {\em a closed area} every time. It is correct that the above setup yields two different signs dependent on how one closes the circles by measurement. But what is the induced voltage if we theoretically close the wire parallely to the wire with no area covered?

The explanation so far seems to lead to the conclusion that the voltage in an open wire by itself remains undetermined until we close the loop and can apply the Faraday law. This is fortunately not the case. In fact even an open wire possesses an induced voltage at its ends if it rests in a time-varying magnetic field. This is due to the fact that a homogeneous magnetic field can be only realized in an asymptotic limit of a finite geometry. This implies certain symmetry points or lines fixing an origin of the coordinate system. We will show that this unavoidably leads to an induced voltage which is the area spanned by the open wire with respect to this symmetry point or line of the magnetic-field-creating setup.

\subsubsection{Plan of the paper}
The argumentation we want to support first by a general explicit calculation from Maxwell equations and we will illustrate it with the help of two exactly solvable models where the homogeneous and time-varying magnetic field is realized in an asymptotic way. First we derive the general formula for the induced voltage showing that the transverse and longitudinal parts contribute with $2/3$ and $1/3$ respectively. Then in section III we present some exactly solvable models which illustrate the necessity to consider the geometry of the source of the magnetic field providing a unique induced voltage. The summary contains a suggestion for determining these geometries and the appendix presents four different ways to calculate a used integral. 

\section{Induction in open wires}

First we consider the general formulas for the induction.
Doubtlessly we can find the induced voltage in a wire if we integrate the present electric fields along the wire
\be
U^{\rm ind}=\int\limits_1^2 \V E\cdot d\V r.
\ee
The question is only which electric fields are present. Therefore we try to find an answer by solving the Maxwell equations. The first equation we consider,
\be 
\V \nabla \times \V E=-\dot {\V B},
\ee 
is best Fourier transformed by
\be
\V E_k(\omega)=\int d^3 r {\rm e}^{i\V k\cdot \V r}\int d t{\rm e}^{-i\omega t} E(\V r,t)
\ee
to take the form
\be
\V k\times \V E_k=i \dot {\V {B_k}}.
\label{1}
\ee
Observing the second Maxwell equation that there are no magnetic monopoles, $\V \nabla\cdot\V B=0$, or Fourier transformed $\V k\cdot \V B_k=0$, the equation (\ref{1}) is solved by vector algebra as
$\V E_k =\V E_k^{\rm trans}+\V E_k^{\rm long}$ where the transverse field reads
\be
\V E_k^{\rm trans}={i\over k^2}\dot {\V B}_k\times \V k.
\label{trans}
\ee
The longitudinal field given by a divergence of a source is not determined by (\ref{1}) since $\V \nabla\times \V \nabla \rho=0$. In order to determine this longitudinal field we must employ Gau\ss{} law as third Maxwell equation
\be
{\rm div} \V E={\rho\over \epsilon\epsilon_0}
\label{2}
\ee
with the vacuum permittivity $\epsilon_0$ and the dielectric constant $\epsilon$ characterizing the medium. In contrast to (\ref{1}) which determines only the transverse electric field unambiguously, the Gau\ss{} law determines only the longitudinal field which reads after Fourier transform
\be
\V E_k^{\rm long}=-{i\rho_k\over \epsilon\epsilon_0k^2} \V k.
\label{long}
\ee

Here we assume a real or virtual charge density $\rho$. It drops out of the final formula but is needed to show that this longitudinal field contributes with one third to the induced voltage.
%
The induced voltage is the line integral along the wire curve $C$ running from $\vec r_1$ to $\vec r_2$. 

Let us emphasize again that the Maxwell equation (\ref{1}) alone does not determine the electric field. One needs additional boundary conditions or the second Maxwell equation (\ref{2}) to determine the complete electric field.

Let us inspect the transverse and longitudinal parts separately. The transverse field is obtained from (\ref{trans}) 
\be
U^{\rm ind}_{\rm trans}&=&\int\limits_C d\V r\cdot \V E^{\rm trans}_r
\nonumber\\
&=&i\int\limits_C \!d\V r\cdot\int\!\!{d^3k\over (2\pi)^3}{\rm e}^{i\V k\cdot \V r}\!\int \!\!d^3 r'{\rm e}^{-i\V k\cdot \V r'}\dot {\V B}_{\V r'}\,\,\times {\V k\over k^2}
\nonumber\\
&=&
{1\over 4 \pi}\int\limits_C \int d^3 r' \dot {\V B}_{\V r'}\cdot \left (\V \nabla_r{1\over |\V r-\V r'|}\,\,\times d\V r\right )
\label{14}
\ee
where we have used the inverse Fourier transform of the Coulomb potential
\be
\int{d^3 k\over (2\pi)^3} {\V k\over k^2}{\rm e}^{i\V k\cdot \V a}=-{i\over 4 \pi}\V \nabla_a{1\over a}
\label{Coul}
\ee
and a trivial rotation of the vector product.

In the further steps we assume a homogeneous but time-dependent magnetic field such that the integral in (\ref{14})
\be
\int d^3 r' \V \nabla_{r}{1\over |\V r-\V r'|}=-{4\pi\over 3} \V r
\label{integ1}
\ee
can be performed (see appendix) with the final result
\be
U^{\rm ind}_{\rm trans}=- \dot {\V B} \cdot {1\over 3}\int\limits_C \V r
\,\,\times d\V r=- {2\over 3}\dot {\V B} \cdot \V A_c.
\label{15}
\ee
The wire covers an area $A_c$ with respect to some origin given by the used coordinate system. This will be found to be fixed due to the source of the magnetic field. We obtain
just 2/3 of the expected result. 

In other words 1/3 must be contributed by the
longitudinal field. Indeed, we can calculate the induced voltage of the longitudinal field (\ref{long}) as
\be
U^{\rm ind}_{\rm long}&=&\int\limits_C d\V r\cdot \V E^{\rm long}_r
\nonumber\\
&=&-i\int\limits_C \!d\V r\cdot\int\!\!{d^3k\over (2\pi)^3}{\rm e}^{i\V k\cdot
  \V r}\!\int \!\!d^3 r'{\rm e}^{-i\V k\cdot \V r'}\rho_k {\V k\over
  \epsilon \epsilon_0 k^2}
\nonumber\\
&=&
\int\limits_C d\V r \cdot \int d^3 r' (\V \nabla_{\V r'}\cdot
\V E_{r'})\left ({\V \nabla_r \over 4 \pi }{1\over |\V r-\V r'|}\right )
\nonumber\\
&=&
-\int\limits_C d\V r \cdot \V \nabla_{\V r}\int d^3 r'
\V E_{r'}\cdot \left ({\V \nabla_{r'} \over 4 \pi }{1\over |\V r-\V r'|}\right )
\nonumber\\
&=&
\frac 1 3 \int\limits_C d\V r \cdot \V \nabla_{\V r}\int d^3 r'
\V E_{r'}\cdot \V r
=\frac 1 3 \int\limits_C d\V r \cdot
\V E_{r}
\label{14l}
\ee
where we used (\ref{Coul}) from second to third line as well as (\ref{2}). A corresponding partial integration is performed when going 
from the third to the fourth line. Finally from the fourth to the fifth line we have used (\ref{integ1}).

Since $\int_C d\V r \cdot
\V E_{r}$ is supposed to be the total induced voltage, the longitudinal part (\ref{14l}) provides obviously only 1/3 of
the total induced voltage. We obtain the result that the transverse part of the electric field which is the non-conserving part, contributes with $2/3$ and the conserving longitudinal part with $1/3$ to the total induced voltage.

As expected, the longitudinal part is conserving in the sense that it is just the scalar potential difference seen from the third line of (\ref{14l})
\ba
U^{\rm ind}_{\rm long}=
{1\over 4 \pi\epsilon \epsilon_0}\int\limits_C \!d\V r \int \!d^3 r' \rho(\V r')\V \nabla_r{1\over |\V r\!-\!\V r'|}=\Phi(\V r_2)\!-\!\Phi(\V r_1)
\label{l15}
\end{align}
since the potential of a charge distribution itself is given by
\be
\Phi(\V r)={1\over 4 \pi\epsilon \epsilon_0}\int \!d^3 r' {\rho(\V r')\over |\V
  r\!-\!\V r'|}.
\ee
In other words the longitudinal field induces a voltage which is given by the
difference of the potentials at the ends of the wire but represents only 1/3
of the total induced voltage. Understanding the total induced voltage as the electromotive force we see that the latter is every-time larger than the potential difference as stated in \cite{Gr12}.

\begin{figure}[]
\includegraphics[width=3.5cm]{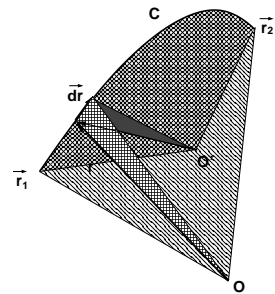}
\caption{Area spanned by the open wire according to (\ref{ar}) with respect to two different origins
  $O$ and $O'$ of coordinates.}
\label{area}
\end{figure}

Summarizing we obtain the induced voltage of an open wire in a homogeneous
time-varying magnetic field as
\be
U^{\rm ind}=- \dot {\V B} \cdot \V A_c
\label{ind}
\ee
with the area spanned by the wire
\be
\V A_c= \frac 1 2 \int\limits_C \V r
\,\,\times d\V r.
\label{ar}
\ee
The formula (\ref{ind}) provides the induced current of a curved wire in a
homogeneous time-dependent magnetic field. The amazing feature
of (\ref{ind}) is now that for an open wire its value depends on
the point of origin $O$ from which we count the curve $C$ as illustrated in figure
\ref{area}. Only in closed curves this point of coordinate origin is dropping
out of the integral. In other words the
value for the induced voltage in open wires is given by the origin of the coordinate system of the geometry which realizes 
the homogeneous magnetic field. One should note that the occurring electric fields can be made visible by the classroom demonstration of induced non-conservative electric fields \cite{vD05}.

\section{Exactly solvable models for asymptotic geometry}

\subsection{General formulas for induced voltage in open wires}

 In order to convince ourselves about the 
above statement that the induced voltage of an open wire is dependent on the
symmetry point of the creating magnetic field, we construct
 a time-varying homogeneous magnetic field by the explicit solution of the
 Maxwell equations.

This solution is conveniently given by the Li\'enard Wiechert potentials \cite{J99}. The divergence-free magnetic field is represented by the vector potential $\V B=\V \nabla \times \V A$ and the Maxwell equation $\dot {\V B}=-\nabla \times \V E$ leads to the relation $\V \nabla \times (\dot {\V A}+\V E)=0$ which means that the electric field is given in terms of the vector potential and a scalar potential $\Phi$
\be
\V E=-\dot {\V A} -\V \nabla \Phi.
\label{e}
\ee
With the Gau\ss{} law it leads to
\be
{\rho \over \epsilon\epsilon_0}=\V \nabla \V E=-\V \nabla \dot {\V A}-\nabla^2\Phi.
\label{1g}
\ee
Using the other Maxwell equation $\mu \mu_0(\V j+\epsilon\epsilon_0 \dot {\V E})=\V \nabla
\times \V B=\V \nabla (\V \nabla \V A)-\V \nabla^2\V A$ one has
\be
\nabla^2 \V A-{1\over c^2}\ddot {\V A}=-\mu \mu_0 \V j+\V \nabla ({1\over c^2} \dot \Phi+\V \nabla \V A).
\label{2g}
\ee
Choosing the Lorenz-gauge ${1\over c^2} \dot \Phi+\V \nabla \cdot \V A=0$, from (\ref{1g}) and (\ref{2g}) the symmetric wave equations for the vector and scalar potential
results
\be
\nabla^2 A-{1\over c^2} \ddot {\V A}&=&-\mu \mu_0\V j\nonumber\\
\nabla^2\Phi-{1\over c^2} \ddot \Phi&=&-{\rho\over \epsilon\epsilon_0}.
\label{wave}
\ee
These inhomogeneous wave equations are solved by retarded potentials
\be
\V A(\V r,t)&=&{\mu\mu_0\over 4 \pi} \int {\V j \left (\V r', t-{|\V r-\V r'|\over
      c}\right )\over |\V r-\V r'|} d^3r'
\nonumber\\ 
\Phi(\V r,t)&=&{1\over 4 \pi \epsilon\epsilon_0} \int {\rho \left (\V r', t-{|\V r-\V r'|\over c}\right )\over |\V r-\V r'|} d^3r' 
\label{aa}
\ee
with the current density $\V j(\V r,t)$ and the charge density $\rho(\V r,t)$. 
Employing the retarded potentials we ensure the consistency with all Maxwell equation. 

Now it is advantageous to express the induced voltage in terms of the vector
potential. The induced voltage is given as a line integral along the wire over
the electric field present at the corresponding point.
Instead of calculating the induced voltage along a curve ${\cal C}: \V r=\V r(q)$ with $q_1\le q\le q_2$, it is more convenient in the following to calculate their time derivative with the help of the Maxwell equation 
\ba
\dot U^{\rm ind}=\int\limits_{\cal C} \dot {\V E} d\V r=-\mu \mu_0 c^2 \int\limits_{\cal C} {\V j} d\V r+c^2 \int\limits_{\cal C}  \V \nabla \times {\V B} d\V r. 
\label{inter0}
\end{align}
The second integral becomes with the help of $\V \nabla\times(\V \nabla\times \V
A)=\V \nabla (\V \nabla \cdot \V A)-\V \nabla^2\V A$
\ba
&c^2\int\limits_{q_1}^{q_2} d q (\V \nabla \cdot \V A)-c^2 \int\limits_{\cal C} \V \nabla^2 \V A
d\V r\nonumber\\
&=\left . c^2 \V \nabla\cdot \V A\right |_{q_1}^{q_2}-\int\limits_{\cal C} d\V r \ddot{\V
  A}+c^2\mu \mu_0\int\limits_{\cal C} d\V r \cdot \V j 
\label{inter}
\end{align}
where we used the wave equation (\ref{wave}) for the vector potential in the
last line. One see that the last part of (\ref{inter}) cancels just the first part of (\ref{inter0}) and we obtain
\be
\dot U^{\rm ind}=-\int\limits_{\cal C} d\V r \ddot {\V A}+c^2\left . \V \nabla\cdot \V A\right |_{q_1}^{q_2}.
\label{open}
\ee
Due to the Lorenz gauge the last part is nothing but the difference of the potential at the ends of the wire. This part vanishes for a closed loop $q_1=q_2$ and one obtains exactly the time derivative of the Faraday law
\ba
\dot U^{\rm ind}=-\oint d \V r \ddot {\V A}&=-\iint d\V F \V\nabla \times \ddot A
=-\iint \ddot {\V B} d \V F 
\end{align}

For an open loop the formula (\ref{open}) is convenient to use since the vector potential is given in terms of the magnetic-field-creating currents which provides a unique result if we integrate over the parameter range $q_1<q<q_2$ describing the wire.

\subsection{Simple parametrization}
One may simply use the representation of the homogeneous magnetic field
\be
\V A=B(t) (z,0,0);\qquad \V B=B(t) (0,1,0)
\label{z}
\ee
Since $\nabla^2 \V A=0$ we have from (\ref{2g}) 
\be
\ddot{\V A}={\V j\over \epsilon \epsilon_0}=\ddot B (z,0,0)
\ee
and $\dot \Phi=0$ or $\rho=0$ which identifies the current as source of this homogeneous and time-varying magnetic field. From (\ref{open}) one gets now for the time derivative of the induced voltage
\be
\dot U^{\rm ind}=-\ddot B \int\limits_C dx z(x)
\label{new}
\ee
which means that it is determined by the area formed by the wire with the $x$-axis perpendicular to the magnetic field direction which is the $y$-axes. In other words, we have a symmetry line, the x-axes, with respect to which the area has to be calculated. Now we will see how a homogeneous magnetic field is asymptotically realized by a finite setup of geometry. 

\subsection{Infinite cylindrical coil}

\begin{figure}[]
\includegraphics[width=4cm]{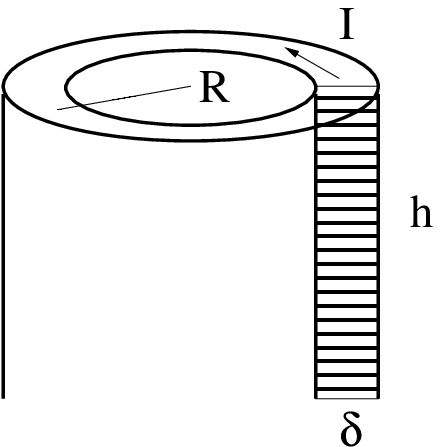}
\vspace{0.5cm}
\includegraphics[width=7.cm]{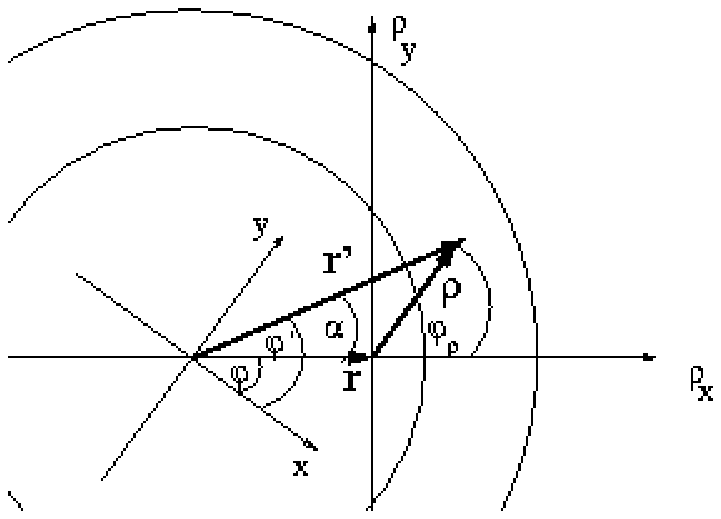}
\caption{Considered geometry of an infinite coil with the current density
  $j=I(t)/h\delta$ running through the shaded area (left) and the chosen integration path (right).}
\label{cyl}
\end{figure}

Let us consider an infinite cylindrical coil with height $h\to\infty$ and an
inner and outer radius of $r_{1/2}=R\pm \delta/2$ as illustrated in figure
\ref{cyl} with a current $I(t)=I_0\cos{\omega t}$ running though an
infinitesimal thick wall.
We transform $\V r'-\V r=\V \rho$ in (\ref{aa}) according to figure
\ref{cyl} and obtain
\ba
\V A={\mu \mu_0 I_0\over 4\pi\delta h}\int \limits_0^ {2\pi} \!\!d \phi_\rho
\!\int\limits_{\rho_-}^{\rho_+}\!\!d \rho \rho \! \int\limits_{\frac h 2}^
{\frac h 2} \!\!\!d z'{\cos{\omega(t\!-\!{\sqrt{\rho^2\!+\!(z\!-\!z')^2}\over c})}\over
  \sqrt{\rho^2\!+\!(z\!-\!z')^2}}\V e_{\phi'}
\label{26}
\end{align}
 with $\V e_{\phi'}=(-\sin{\phi'}, \cos{\phi'},0)$. The cos theorem leads to
 the lower and upper integration limits of
 $\rho_{\pm}=\sqrt{(R\pm\delta/2)^2-r^2\sin^2{\phi_\rho}}-r\cos{\phi_\rho}$
 and the sin theorem leads to $\sin{\phi_\rho}={r'\over \rho} \sin{\alpha}$
 with $\alpha=\phi'-\phi$. 

In the following we will work in the limit of large
 $R$. Since 
\be
{R+\frac \delta 2\over \rho_-}\le {r'\over \rho}\le {R-\frac
     \delta 2\over \rho_+}
\ee 
one finds $r'=\rho+o(r/R)$ and $\phi'=\phi+\phi_\rho+o(r/R)$. We substitute $p=
\sqrt{\rho^2\!+\!(z\!-\!z')^2}$ in (\ref{26}) which allows to perform this integration with the upper and lower limits
$p_\pm=R-r \cos{\phi_\rho}\pm \frac \delta 2 +o(r/R)$ and obtain 
\ba
\V A&=\!-{\mu \mu_0 I_0 c\over 2\pi\omega}\!\!\int \limits_0^{2\pi} \!\!d \!\phi_\rho
\cos{\left [\omega \left (t\!-\!{R\over c}\right )\!+\!{\omega r\over c}\cos{\phi_\rho}\right ]}{\sin{\!\left (\!{\omega \delta\over 2 c}\!\right )}\over \delta}\V e_{\phi'}
\nonumber\\
&=-{\mu \mu_0 I_0 \over 4\pi}\int \limits_0^{2\pi} \!\!\!d\! \phi_\rho
\cos{\left [\omega t'\!+\!{\omega r\over c}\cos{\phi_\rho}\right ]}\begin{pmatrix}-\sin{(\phi\!+\!\phi_\rho)}\cr \cos{(\phi\!+\!\phi_\rho)}\cr 0\end{pmatrix} .
\end{align}
The integration $z'$ over the length $h$ has become trivial and in the last line we have performed the limit $\delta\to 0$. Further we see that the time is delayed by $t'=\left (t-{R\over c}\right )$ which is the time the field needs to overcome the distance $R$ from the source. The last angular integration can be performed with the final result
\ba
\V A&=-{\mu \mu_0 I_0 \over 2}J_1\left ({\omega r\over c}\right )\begin{pmatrix}-\sin{\phi}\cr \cos{\phi}\cr 0\end{pmatrix}\sin\omega t' 
\label{azyl}
\end{align}
with the Bessel function 
$J_1(x)=x[J_0-J_1'(x)]$.
The magnetic field becomes
\be
\V B=\V \nabla \times \V A={\mu \mu_0 I_0 \omega\over 2 c}\sin{\omega t'} \begin{pmatrix}0\cr 0\cr J_0\left ({\omega r\over c}\right )\end{pmatrix} 
\ee
which is directed along the symmetry axis. We see that a homogeneous magnetic field is only realizable for distances $r<<c/\omega$ from the symmetry axes since then $J_0(x)\approx 1$. Within this limit we have $J_1(x)\approx x/2$ and introducing (\ref{azyl}) into  (\ref{open}) we obtain
\ba
\dot U^{\rm ind}={\mu \mu_0 I_0\omega^3\over 4 c}\!\sin\omega t'\!\!\int\limits_C \!\!d\V r  r  \!\begin{pmatrix}\!-\sin{\phi}\cr \cos{\phi}\cr 0\end{pmatrix} 
= \!-\!\ddot B_z \frac 1 2 \int\limits_{\phi_1 }^{\phi_2}\!\!d\phi r(\phi)^2
\end{align}
where we have represented the wire parametrically by $r(\phi)$. We obtain just the time derivative of the induced voltage (\ref{ind}) with the sectoral area (\ref{ar}) spanned by the wire with respect to the $z$-axes. The latter is the symmetry axes provided by the setup of the asymptotic homogeneous magnetic field.  Please note that $\V \nabla\cdot \V A=0$ is exactly valid for (\ref{azyl}), i.e. there are no scalar fields.

We see that the induced voltage in an open wire is just the sector area spanned by the wire with respect to the origin perpendicular to the magnetic field. This origin is given by the $z$-axes as the symmetry axes of the magnetic-field-creating setup.

\subsection{Infinite cylindrical plates}
\begin{figure}[]
\includegraphics[width=5cm]{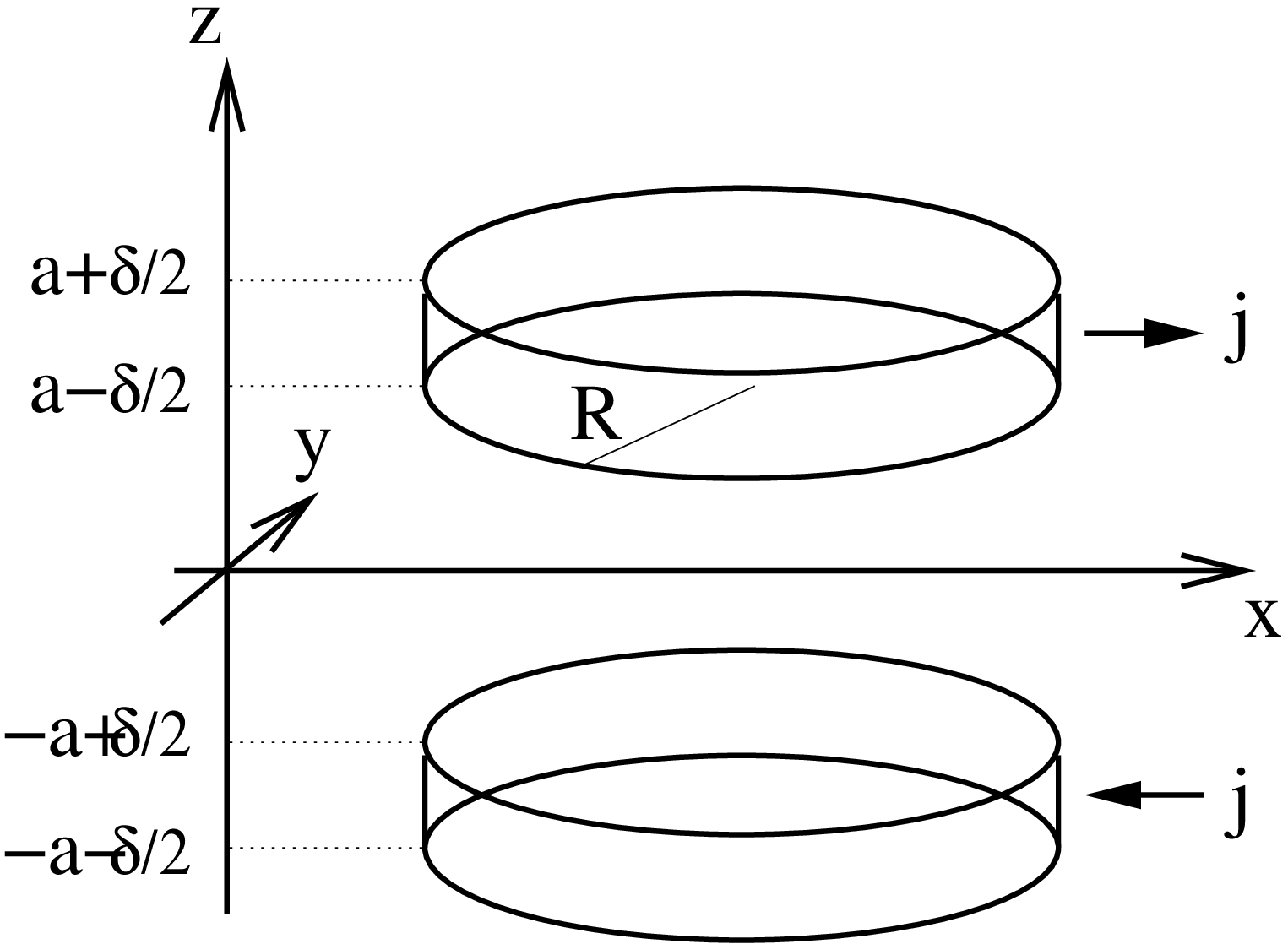}
\vspace{0.3cm}
\includegraphics[width=6cm]{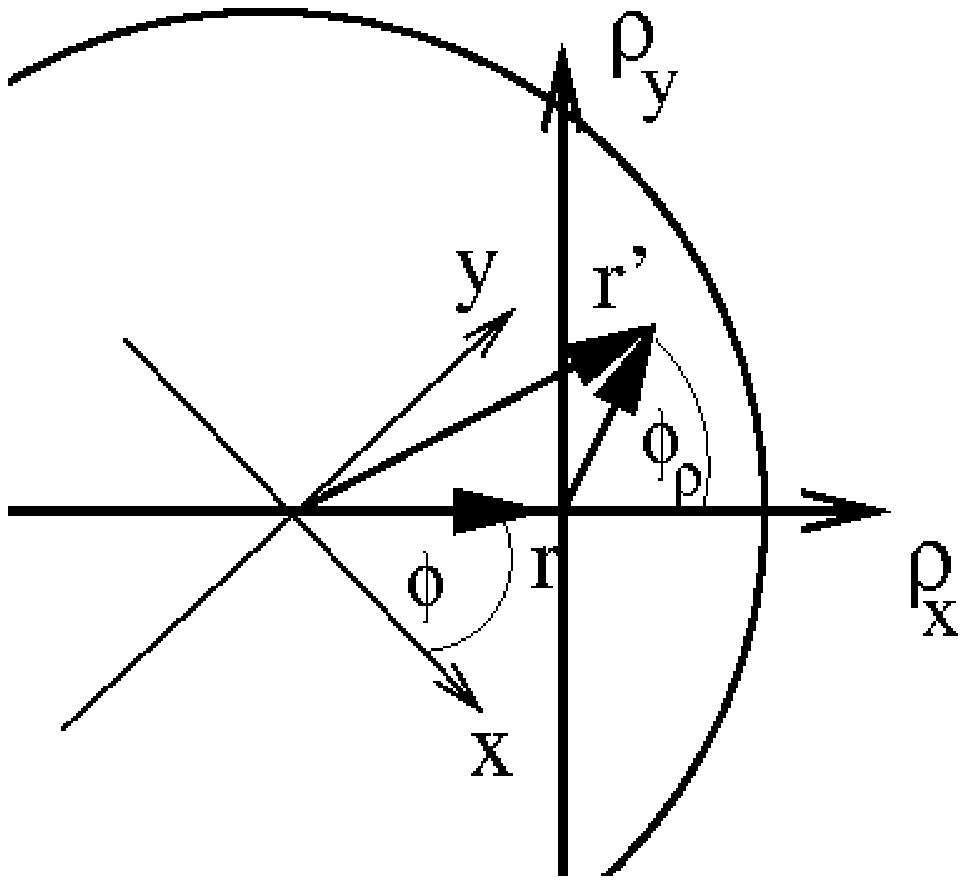}
\caption{Considered geometry of two antiparalelly running currents through cylindrical plates at the distances $\pm a$ from the x-axes (left). The integration variables for perpendicular vectors $\V \rho={\bf r'}-{\bf r}$ are seen at the right side.}
\label{integ}
\end{figure}

In a second example we want to consider a situation where we do not have a symmetry point but a symmetry line. 
We construct the time-varying homogeneous magnetic field by two time-varying currents which run in x-direction in an upper plate at $(0,0,a)$ and in opposite direction in a plate at $(0,0,-a)$. We assume that the plates are cylinders with the thickness $\delta$ in $z$-direction and a radius $R$ in $x$ and $y$-direction as illustrated in figure \ref{integ}. We will consider the limit of small $\delta$ and large $R$ which should produce a homogeneous and time-varying magnetic field in $y$-direction near the center for small $z$. 

The current runs through the area $2 R \delta$. With the current per area $I/2 R=j_0\cos\omega t$ we can define the line current density for the upper and lower plate as
\be
\V j_\pm=\pm \V e_x {I(t)\over 2 R \delta}=\pm \V e_x {j_0\cos{\omega t}\over \delta}.
\ee
We consider the line current density with respect to the line $2R$ seen by current $I$. Since we will work in the limit $R\to \infty$ the difference between lateral midpoint and endpoints does not matter.
  
The vector potential follows the current to be in x-direction $\V A=(0,0,A_a-A_{-a})$ and reads in cylinder coordinates from figure \ref{integ}
\ba
A_a={\mu \mu_0 j_0 \over 4 \pi \delta}\!\int\limits_0^{2\pi} d\phi_\rho \! \int\limits_0^{\rho_m} d \rho \rho
\!\!\int\limits_{a-\delta/2}^{a+\delta/2}
\!\!\!d z'\,{\cos{\!\left (\!t\!-\!{\sqrt{\rho^2\!+\!(z\!-\!z')^2}\over c}\right )}\over \sqrt{\rho^2\!+\!(z\!-\!z')^2}}   
\end{align}
for the part from the upper plate. The part of the lower plate has to be subtracted with the replacement $a\to -a$. 

For the upper limit  one gets $\rho_m=\sqrt{R^2-r^2\sin^2\phi_\rho}-r\cos\phi_\rho$. With the limit of small $\delta$
\be
{1\over \delta}\int\limits_{a-\delta/2}^{a+\delta/2} \!\!d z'\, f(z')=f(a)+o(\delta^{-1}),
\ee
and replacing $p=\sqrt{\rho^2+(z-a)^2}$,
the $\rho$ integration yields
\be
A_a&=&{\mu \mu_0 j_0 c\over 4 \pi \omega} \int\limits_0^{2\pi} d\phi_\rho 
\left [
\sin{\omega \left (t-{|a-z|\over c}\right )}
\right.\nonumber\\ && \left .
-
\sin{\omega \left (t-{\sqrt{\rho_m^2+(a-z)^2}\over c}\right )}
\right ].
\label{34}
\ee
We can now calculate all quantities explicitly and use the limit $R\to \infty$ 
which provides $\sqrt{\rho_m^2+(a-z)^2}=R-r\cos \phi_\rho+o(r/R)$ and the second $sin$ term in (\ref{34}) is subtracted when the contribution of both plates are added. One obtains
\ba
A_x=A_a-A_{-a}={\mu \mu_0 c j_0\over \omega} \cos{\left [\omega \left (t-{a\over c}\right )\right ]} \sin{\left [{\omega z\over c}\right ]}
\label{a0}
\end{align}
and the magnetic field becomes 
\be
\V B=\left \{0,\mu \mu_0 j_0 \cos{\omega \left (t-{a\over c}\right )}\cos{\omega z\over c},0\right \}.
\ee
We see again the appearance of time delay $t'=t-{a\over c}$ which is the time the field needs to overcome the distance from the source current. A homogeneous magnetic field is only asymptotically possible for $z\ll c/\omega$.

In this limit $o({w z/ c})^2$ we obtain for the time derivative of the induced voltage (\ref{open})
\ba
\dot U^{\rm ind}=\mu \mu_0 j_0 \omega^2 \cos {\omega t'}\!\!\int\limits_{q_1}^{q_2} \!\!d q {d x(q)\over d q}z(q)
=-\ddot B_y\!\!\int\limits_{x_1}^{x_2} \!\!d x z
\label{final}
\end{align}
which is (\ref{new}) and has the general form (\ref{ind}). The area, however, is now the one which the wire forms with the x-axes perpendicular to the magnetic field direction which is the y-axes. For a closed wire the last integral gives again the Faraday law for the enclosed area in $x,y$ direction which is penetrated by the time-dependent magnetic field. For an open wire we see that the area is now given by a symmetry line, the x-axes, and not the symmetry point as in the last example. Please not again that $\V \nabla \cdot \V A=0$ for (\ref{a0}) which means that no scalar potentials are present.

\section{Summary}
Irrespective of the actual procedure to realize a homogeneous and time-varying
magnetic field one can integrate the time derivative of the electric field
along the line to obtain the time derivative of the induced voltage. The longitudinal electric field provides only
1/3 of the induced voltage and is given by the potential difference
between the ends of the wire. The transverse and non-conserving field contributes with 2/3 to the induced voltage.
 A homogeneous magnetic field can be only realized in an asymptotic limit of a fixed geometrical setup. The latter one defines certain symmetry axes or symmetry points and the origin of the coordinate system. We find from the general solution of the Maxwell equation that the induced voltage of an open wire is given by the area spanned with respect to this symmetry point or line. Correspondingly we obtain a sector formula for the area if a cylindrical field is present and an area integral with respect to a line in a planar symmetry. We have illustrated these two cases by two exactly
solvable models for the Maxwell equations realizing the homogeneous and
time-varying magnetic field. 

The dependence of the open circuit voltage on the direction of the magnetic field has been measured in \cite{DCGL05}. There it has been found that if the surface of the electrode is oriented parallel
or perpendicular to the magnetic field, the open circuit potential moves in opposite directions with the largest changes occurring when the electrode surface is parallel to the
magnetic field. This observation is explained by our findings.

If the wire crosses the symmetry line or point we can establish a simple mirror rule by considering the exact mirror image of the wire in a perpendicular plane to the magnetic field.
The wire and its image should have the same induced voltage if the wire crosses the symmetry line. Therefore if we close the open wire with its image we should have twice the induced voltage of the open wire. The now closed area obeys Faraday's law. Therefore we can suggest the rule that the induced voltage in this case is half the one which is induced by the area covered by the wire and its mirror image. Alternatively we might connect the curved wire with a straight line and use this area for the induction law. This mirror rule is only applicable if the symmetry axes or point crosses the wire.


We can suggest an experimental setup to determine the symmetry point or symmetry line of a given magnetic field. Measuring the induced voltage of an open straight line wire in different directions would yield zero if the wire is aligned perpendicular to a symmetry axes. If this line wire is now rotated from $0^o$ to $90^o$ one can extract from the increasing voltage the geometry whether we have sector formula as in our first example or a area integral with respect to a symmetry axes. From this one can conclude about the origin of the asymptotically homogeneous magnetic field. This might have astrophysical applications in determining the symmetry of sources of magnetic fields.

\acknowledgments
The authors are grateful to Martin Poppe about questioning the paradox.

\appendix

\section{Four ways to calculate an integral}
We are going to calculate the integral
\be
\V I=\int d^3r'\V \nabla_{r} {1\over |\V r-\V r'|}=-{4 \pi\over 3} \V r
\label{int}
\ee
in four different ways. 

\paragraph{By Gau\ss{}-Ostrogatzky}
The integral can be directly transformed into a surface integral by the
integral theorem of Gau\ss{}-Ostrogatzky (one writes Gau\ss{} theorem three times for each coordinate and combines it as a vector) 
\be
\int d^3 r' \V \nabla_{r'} g=\oiint d \V A \,g
\ee
and performing the azimuthal angle integration leaving the altitudes, one gets
\be
\V I&=&-\oiint {1\over |\V r\!-\!\V r'|} d\V A'=2 \pi\!\!\lim\limits_{r'\to \infty} \!r'^2{\V r\over r} \int\limits_{-1}^1d x {x\over \sqrt{r^2\!+\!r'^2\!-\!2 r r' x}}
\nonumber\\
&=&-\lim\limits_{r'\to \infty}{4 \pi\over 3}{\V r}\left \{ \begin{matrix}1&r'>r\cr
{r'^3\over r^3} &r'<r\end{matrix} \right .=-{4 \pi\over 3}{\V r}
\ee

\paragraph{Direct integration}

Performing the azimuthal angle integration directly
\be
\V I&=&\int d^3 r'{\V r'\!-\!\V r\over |\V r\!-\!\V r'|^3}=2 \pi \V r \int\limits_0^\infty d y y^2 \int\limits_{-1}^1 dx {y x\!-\!1\over (1\!+\!y^2\!-\!2 y x )^{3/2}}
\nonumber\\
&=&-{4 \pi\over 3}\V r
\label{direct}
\ee

\paragraph{By limit of known integrals}

The known mean Coulomb energy in s-wave state leads to the integral
\ba
\int d^3r {{\rm e}^{-2 r/a}\over |\V R-\V r|}=\pi a^3 \left [{1\over R}(1-{\rm e}^{-2R/a})-{1\over a}{\rm e}^{-2R/a}\right ]
\end{align}
which we can use to apply $-\V \nabla_R$ and performing the $a\to\infty$ limits
leads exactly to (\ref{int}).

\paragraph{By a vector trick}
Using $\V \nabla_{r'} \cdot  \V r'=3$ we can use partial integration for the i-th component
\be
I_i&=&-\frac 1 3 \int d^3r'(\V \nabla_{r'}\cdot \V r') \partial_i'{1\over |\V r-\V r'|}
\nonumber\\&=&\frac 1 3 \int d^3r' r_j'\partial_i'\partial_j'{1\over |\V r-\V r'|}=-{4\pi \over 3}\delta_{ij}\int d^3 r' r_j'\delta (\V r-\V r')
\nonumber\\
&=&-{4\pi \over 3} r_i
\ee
since all non-diagonal combinations are zero due to angular integrations and we
have used
$
\V \nabla^2_r{1\over |\V r-\V r'|}= -{4\pi}\delta (\V r-\V r')
$.

\bibliography{bose,kmsr,kmsr1,kmsr2,kmsr3,kmsr4,kmsr5,kmsr6,kmsr7,delay2,spin,spin1,refer,delay3,gdr,chaos,sem3,sem1,sem2,short,cauchy,genn,paradox,deform,shuttling,blase}

\begin{thebibliography}{17}%
\makeatletter
\providecommand \@ifxundefined [1]{%
 \@ifx{#1\undefined}
}%
\providecommand \@ifnum [1]{%
 \ifnum #1\expandafter \@firstoftwo
 \else \expandafter \@secondoftwo
 \fi
}%
\providecommand \@ifx [1]{%
 \ifx #1\expandafter \@firstoftwo
 \else \expandafter \@secondoftwo
 \fi
}%
\providecommand \natexlab [1]{#1}%
\providecommand \enquote  [1]{``#1''}%
\providecommand \bibnamefont  [1]{#1}%
\providecommand \bibfnamefont [1]{#1}%
\providecommand \citenamefont [1]{#1}%
\providecommand \href@noop [0]{\@secondoftwo}%
\providecommand \href [0]{\begingroup \@sanitize@url \@href}%
\providecommand \@href[1]{\@@startlink{#1}\@@href}%
\providecommand \@@href[1]{\endgroup#1\@@endlink}%
\providecommand \@sanitize@url [0]{\catcode `\\12\catcode `\$12\catcode
  `\&12\catcode `\#12\catcode `\^12\catcode `\_12\catcode `\%12\relax}%
\providecommand \@@startlink[1]{}%
\providecommand \@@endlink[0]{}%
\providecommand \url  [0]{\begingroup\@sanitize@url \@url }%
\providecommand \@url [1]{\endgroup\@href {#1}{\urlprefix }}%
\providecommand \urlprefix  [0]{URL }%
\providecommand \Eprint [0]{\href }%
\providecommand \doibase [0]{http://dx.doi.org/}%
\providecommand \selectlanguage [0]{\@gobble}%
\providecommand \bibinfo  [0]{\@secondoftwo}%
\providecommand \bibfield  [0]{\@secondoftwo}%
\providecommand \translation [1]{[#1]}%
\providecommand \BibitemOpen [0]{}%
\providecommand \bibitemStop [0]{}%
\providecommand \bibitemNoStop [0]{.\EOS\space}%
\providecommand \EOS [0]{\spacefactor3000\relax}%
\providecommand \BibitemShut  [1]{\csname bibitem#1\endcsname}%
\let\auto@bib@innerbib\@empty
\bibitem [{\citenamefont {Galili}\ \emph {et~al.}(2006)\citenamefont {Galili},
  \citenamefont {Kaplan},\ and\ \citenamefont {Lehavi}}]{GIKLY06}%
  \BibitemOpen
  \bibfield  {author} {\bibinfo {author} {\bibfnamefont {I.}~\bibnamefont
  {Galili}}, \bibinfo {author} {\bibfnamefont {D.}~\bibnamefont {Kaplan}}, \
  and\ \bibinfo {author} {\bibfnamefont {Y.}~\bibnamefont {Lehavi}},\
  }\href@noop {} {\bibfield  {journal} {\bibinfo  {journal} {American Journal
  of Physics}\ }\textbf {\bibinfo {volume} {74}},\ \bibinfo {pages} {337}
  (\bibinfo {year} {2006})}\BibitemShut {NoStop}%
\bibitem [{\citenamefont {Griffiths}(2013)}]{Gr12}%
  \BibitemOpen
  \bibfield  {author} {\bibinfo {author} {\bibfnamefont {D.~J.}\ \bibnamefont
  {Griffiths}},\ }\href@noop {} {\emph {\bibinfo {title} {Introduction to
  Electrodynamics}}}\ (\bibinfo  {publisher} {Addison-Wesley},\ \bibinfo
  {address} {Cloth},\ \bibinfo {year} {2013})\BibitemShut {NoStop}%
\bibitem [{\citenamefont {Scorgie}(1995)}]{Sco95}%
  \BibitemOpen
  \bibfield  {author} {\bibinfo {author} {\bibfnamefont {G.~C.}\ \bibnamefont
  {Scorgie}},\ }\href@noop {} {\bibfield  {journal} {\bibinfo  {journal}
  {European Journal of Physics}\ }\textbf {\bibinfo {volume} {16}},\ \bibinfo
  {pages} {36} (\bibinfo {year} {1995})}\BibitemShut {NoStop}%
\bibitem [{\citenamefont {Kim}\ \emph {et~al.}(2008)\citenamefont {Kim},
  \citenamefont {Hwang}, \citenamefont {Hwang},\ and\ \citenamefont
  {Ahn}}]{KHHA08}%
  \BibitemOpen
  \bibfield  {author} {\bibinfo {author} {\bibfnamefont {H.~K.}\ \bibnamefont
  {Kim}}, \bibinfo {author} {\bibfnamefont {J.~S.}\ \bibnamefont {Hwang}},
  \bibinfo {author} {\bibfnamefont {S.-W.}\ \bibnamefont {Hwang}}, \ and\
  \bibinfo {author} {\bibfnamefont {D.}~\bibnamefont {Ahn}},\ }\href {\doibase
  10.1109/TNANO.2007.909068} {\bibfield  {journal} {\bibinfo  {journal}
  {Nanotechnology, IEEE Transactions on}\ }\textbf {\bibinfo {volume} {7}},\
  \bibinfo {pages} {120} (\bibinfo {year} {2008})}\BibitemShut {NoStop}%
\bibitem [{\citenamefont {Khanna}\ and\ \citenamefont {LeBlanc}(1972)}]{KL72}%
  \BibitemOpen
  \bibfield  {author} {\bibinfo {author} {\bibfnamefont {S.~M.}\ \bibnamefont
  {Khanna}}\ and\ \bibinfo {author} {\bibfnamefont {M.~A.~R.}\ \bibnamefont
  {LeBlanc}},\ }\href@noop {} {\bibfield  {journal} {\bibinfo  {journal} {J.
  Appl. Phys.}\ }\textbf {\bibinfo {volume} {43}},\ \bibinfo {pages} {5165}
  (\bibinfo {year} {1972})}\BibitemShut {NoStop}%
\bibitem [{\citenamefont {Levin}\ and\ \citenamefont {Rizzato}(2006)}]{LR06}%
  \BibitemOpen
  \bibfield  {author} {\bibinfo {author} {\bibfnamefont {Y.}~\bibnamefont
  {Levin}}\ and\ \bibinfo {author} {\bibfnamefont {F.~B.}\ \bibnamefont
  {Rizzato}},\ }\href@noop {} {\bibfield  {journal} {\bibinfo  {journal} {Phys.
  Rev. E}\ }\textbf {\bibinfo {volume} {74}},\ \bibinfo {pages} {066605}
  (\bibinfo {year} {2006})}\BibitemShut {NoStop}%
\bibitem [{\citenamefont {Lipavsk{\'y}}\ \emph {et~al.}(2007)\citenamefont
  {Lipavsk{\'y}}, \citenamefont {Kol{\'a}{\v c}ek}, \citenamefont {Morawetz},
  \citenamefont {Brandt},\ and\ \citenamefont {Yang}}]{LKMBY07}%
  \BibitemOpen
  \bibfield  {author} {\bibinfo {author} {\bibfnamefont {P.}~\bibnamefont
  {Lipavsk{\'y}}}, \bibinfo {author} {\bibfnamefont {J.}~\bibnamefont
  {Kol{\'a}{\v c}ek}}, \bibinfo {author} {\bibfnamefont {K.}~\bibnamefont
  {Morawetz}}, \bibinfo {author} {\bibfnamefont {E.~H.}\ \bibnamefont
  {Brandt}}, \ and\ \bibinfo {author} {\bibfnamefont {T.~J.}\ \bibnamefont
  {Yang}},\ }\href@noop {} {\emph {\bibinfo {title} {Bernoulli potential in
  superconductors}}}\ (\bibinfo  {publisher} {Springer},\ \bibinfo {address}
  {Berlin},\ \bibinfo {year} {2007})\ \bibinfo {note} {{L}ecture Notes in
  Physics 733, ISBN 978-3-540-73455-0}\BibitemShut {NoStop}%
\bibitem [{\citenamefont {Spavieri}(2012)}]{Sp12}%
  \BibitemOpen
  \bibfield  {author} {\bibinfo {author} {\bibfnamefont {G.}~\bibnamefont
  {Spavieri}},\ }\href@noop {} {\bibfield  {journal} {\bibinfo  {journal} {Eur.
  Phys. J. D}\ }\textbf {\bibinfo {volume} {66}},\ \bibinfo {pages} {76}
  (\bibinfo {year} {2012})}\BibitemShut {NoStop}%
\bibitem [{\citenamefont {Redinz}(2011)}]{Re11}%
  \BibitemOpen
  \bibfield  {author} {\bibinfo {author} {\bibfnamefont {J.~A.}\ \bibnamefont
  {Redinz}},\ }\href@noop {} {\bibfield  {journal} {\bibinfo  {journal}
  {American Journal of Physics}\ }\textbf {\bibinfo {volume} {79}},\ \bibinfo
  {pages} {774} (\bibinfo {year} {2011})}\BibitemShut {NoStop}%
\bibitem [{\citenamefont {Dass}\ \emph {et~al.}(2005)\citenamefont {Dass},
  \citenamefont {Counsil}, \citenamefont {Gao},\ and\ \citenamefont
  {Leventis}}]{DCGL05}%
  \BibitemOpen
  \bibfield  {author} {\bibinfo {author} {\bibfnamefont {A.}~\bibnamefont
  {Dass}}, \bibinfo {author} {\bibfnamefont {J.~A.}\ \bibnamefont {Counsil}},
  \bibinfo {author} {\bibfnamefont {X.}~\bibnamefont {Gao}}, \ and\ \bibinfo
  {author} {\bibfnamefont {N.}~\bibnamefont {Leventis}},\ }\href@noop {}
  {\bibfield  {journal} {\bibinfo  {journal} {J. Phys. Chem. B}\ }\textbf
  {\bibinfo {volume} {109}},\ \bibinfo {pages} {11065} (\bibinfo {year}
  {2005})}\BibitemShut {NoStop}%
\bibitem [{\citenamefont {Waskaas}\ and\ \citenamefont {Kharkats}(2001)}]{W01}%
  \BibitemOpen
  \bibfield  {author} {\bibinfo {author} {\bibfnamefont {M.}~\bibnamefont
  {Waskaas}}\ and\ \bibinfo {author} {\bibfnamefont {Y.~I.}\ \bibnamefont
  {Kharkats}},\ }\href@noop {} {\bibfield  {journal} {\bibinfo  {journal} {J.
  Electroanal. Chem.}\ }\textbf {\bibinfo {volume} {502}},\ \bibinfo {pages}
  {51} (\bibinfo {year} {2001})}\BibitemShut {NoStop}%
\bibitem [{\citenamefont {Perov}\ \emph {et~al.}(2002)\citenamefont {Perov},
  \citenamefont {Sheverdyaeva},\ and\ \citenamefont {Inoue}}]{SI02}%
  \BibitemOpen
  \bibfield  {author} {\bibinfo {author} {\bibfnamefont {N.~S.}\ \bibnamefont
  {Perov}}, \bibinfo {author} {\bibfnamefont {P.~M.}\ \bibnamefont
  {Sheverdyaeva}}, \ and\ \bibinfo {author} {\bibfnamefont {M.}~\bibnamefont
  {Inoue}},\ }\href@noop {} {\bibfield  {journal} {\bibinfo  {journal} {J.
  Appl. Phys.}\ }\textbf {\bibinfo {volume} {91}},\ \bibinfo {pages} {8557}
  (\bibinfo {year} {2002})}\BibitemShut {NoStop}%
\bibitem [{\citenamefont {Sueptitz}\ \emph {et~al.}(2010)\citenamefont
  {Sueptitz}, \citenamefont {Tschulik}, \citenamefont {Uhlemann}, \citenamefont
  {Gebert},\ and\ \citenamefont {Schultz}}]{Su10}%
  \BibitemOpen
  \bibfield  {author} {\bibinfo {author} {\bibfnamefont {R.}~\bibnamefont
  {Sueptitz}}, \bibinfo {author} {\bibfnamefont {K.}~\bibnamefont {Tschulik}},
  \bibinfo {author} {\bibfnamefont {M.}~\bibnamefont {Uhlemann}}, \bibinfo
  {author} {\bibfnamefont {A.}~\bibnamefont {Gebert}}, \ and\ \bibinfo {author}
  {\bibfnamefont {L.}~\bibnamefont {Schultz}},\ }\href@noop {} {\bibfield
  {journal} {\bibinfo  {journal} {Electrochimica Acta}\ }\textbf {\bibinfo
  {volume} {55}},\ \bibinfo {pages} {5200 } (\bibinfo {year}
  {2010})}\BibitemShut {NoStop}%
\bibitem [{\citenamefont {Bund}\ \emph {et~al.}(2005)\citenamefont {Bund}, ,\
  and\ \citenamefont {Kuehnlein}}]{Bu05}%
  \BibitemOpen
  \bibfield  {author} {\bibinfo {author} {\bibfnamefont {A.}~\bibnamefont
  {Bund}}, , \ and\ \bibinfo {author} {\bibfnamefont {H.~H.}\ \bibnamefont
  {Kuehnlein}},\ }\href@noop {} {\bibfield  {journal} {\bibinfo  {journal} {The
  Journal of Physical Chemistry B}\ }\textbf {\bibinfo {volume} {109}},\
  \bibinfo {pages} {19845} (\bibinfo {year} {2005})}\BibitemShut {NoStop}%
\bibitem [{\citenamefont {Dunne}\ and\ \citenamefont {Coey}(2012)}]{Du12}%
  \BibitemOpen
  \bibfield  {author} {\bibinfo {author} {\bibfnamefont {P.}~\bibnamefont
  {Dunne}}\ and\ \bibinfo {author} {\bibfnamefont {J.~M.~D.}\ \bibnamefont
  {Coey}},\ }\href@noop {} {\bibfield  {journal} {\bibinfo  {journal} {Phys.
  Rev. B}\ }\textbf {\bibinfo {volume} {85}},\ \bibinfo {pages} {224411}
  (\bibinfo {year} {2012})}\BibitemShut {NoStop}%
\bibitem [{\citenamefont {Kraftmakher}(2000)}]{Kr99}%
  \BibitemOpen
  \bibfield  {author} {\bibinfo {author} {\bibfnamefont {Y.}~\bibnamefont
  {Kraftmakher}},\ }\href@noop {} {\bibfield  {journal} {\bibinfo  {journal}
  {American Journal of Physics}\ }\textbf {\bibinfo {volume} {68}},\ \bibinfo
  {pages} {375} (\bibinfo {year} {2000})}\BibitemShut {NoStop}%
\bibitem [{\citenamefont {van Deursen}(2005)}]{vD05}%
  \BibitemOpen
  \bibfield  {author} {\bibinfo {author} {\bibfnamefont {A.~P.~J.}\
  \bibnamefont {van Deursen}},\ }\href@noop {} {\bibfield  {journal} {\bibinfo
  {journal} {American Journal of Physics}\ }\textbf {\bibinfo {volume} {73}},\
  \bibinfo {pages} {1099} (\bibinfo {year} {2005})}\BibitemShut {NoStop}%
\end{thebibliography}%

\end{document}